 \documentclass[preprint,showpacs,preprintnumbers,amsmath,amssymb]{revtex4}
\usepackage{amsmath}

\usepackage{graphicx}
\usepackage{dcolumn}
\usepackage{bm}

\begin{document}

\title{Band renormalization and Fermi surface reconstruction in iron-based superconductors}

\author{Shun-Li Yu, Jing Kang, and Jian-Xin Li}
\affiliation{National Laboratory of Solid State Microstructures and
Department of Physics, Nanjing University, Nanjing 210093, China}

\date{\today}

\begin{abstract}
Using the fluctuation exchange approximation and a three-orbital
model, we study the band renormalization, Fermi surface
reconstruction and the superconducting pairing symmetry in the
newly-discovered iron-based superconductors. We find that the
inter-orbital spin fluctuations lead to the strong anisotropic
band renormalization and the renormalization is orbital dependent.
As a result, the topology of Fermi surface displays distinct
variation with doping from the electron type to the hole type,
which is consistent with the recent experiments. This shows that
the Coulomb interactions will have a strong effect on the band
renormalization and the topology of the electron Fermi pocket. In
addition, the pairing state mediated by the inter-orbital spin
fluctuation is of an extended $s$-wave symmetry.
\end{abstract}


\maketitle

\section{INTRODUCTION}

Recently, the discovery of superconductivity in the iron-based
compounds has generated enormous interest, because these materials
are the first non-copper superconductors with high superconducting
(SC) critical temperature. These compounds share the same FeAs(or
FeP) layers that are believed to be responsible for the
superconductivity. Two classes of such compounds have been
extensively investigated: (1) The LaOFeAs classes(denoted as
FeAs-1111), space group $P_{4}/mmm$, with $\mathrm{Tc}\approx26$ K
through electron doping with replacing O$^{2-}$ by
F$^{-}$~\cite{Kamihara} and $T_c$ can be up to about 41 to 56 K by
replacing lanthanum by other rare earth
ions~\cite{ChenXH,ChenGF,Wang}. (2) The BaFe$_{2} $As$_{2}$
classes(denoted as FeAs-122), space group $I_{4}/mmm$, with
$\mathrm{Tc}\approx38$ K through hole doping with substituting
Ba$^{2+}$ for K$^{+}$~\cite{Rotter}. All parent compounds show a
spin-density-wave(SDW) abnormality below a temperature $\sim150$
K~\cite{Clarina,dong,Huang} and the superconductivity is
associated with the suppression of the SDW. Most experiment
measurements have shown that these superconductors open a full gap
around both the hole and electron Fermi
pockets~\cite{zhao,ChenTY,Ding,Richard,Malone,MuG}, though the
NMR~\cite{Zheng,Grafe} and transport~\cite{Gor,Flet} experiments
suggest the presence of gap nodes. At the same time, although the
angle-resolved photoemission spectroscopy(ARPES) measurements of
the band structure and Fermi surface(FS) in the undoped and
electron doped samples~\cite{Lu,LiuC,Yang} are qualitatively in
agreement with the finding of the first-principle band
calculation~\cite{Singh,Nekrasov}, a strong anisotropic band
renormalization is found. On the other hand, the ARPES data for
the hole doped FeAs-122 reveal that the electron FS around
$(\pi,\pi)$ consisting of disconnecting patches~\cite{Liu,zhao},
which exhibits a significant difference from the result obtained
in the band structure calculation. Therefore, the understanding of
the role played by the electron correlation and the origin of the
different electron FS topology between the hole-doped and
electron-doped system is of importance.

As the band structure calculations having shown, the FS and band
structures of these compounds are qualitatively
similar~\cite{Singh,Nekrasov,Boeri} and the Fe-3d orbitals represent
the main contribution to the density of states, together with a
contribution from As-p orbitals, within several eV of the Fermi
level. In the FeAs layers, the Fe atoms form a square lattice and a
Fe atom is coordinated by four As atoms in a tetrahedron. Due to the
direct Fe-Fe bonds and the hybridization with the As-4p orbitals,
the Fe-3d orbitals form a complex band structure. However, the main
contribution to the bands near the Fermi level comes from the
$d_{xz}$, $d_{yz}$ and the $d_{xy}$ orbitals(the direct Fe-Fe bonds
along the $x$ and $y$ axes) of the Fe atoms~\cite{Boeri,Mazin,ZYLu}.
In this paper, we employ a three-orbital (the $d_{xz}$, $d_{yz}$ and
$d_{xy}$ orbitals) model~\cite{Lee} to investigate the band
renormalization, FS reconstruction and superconducting gap symmetry
in the iron-based compounds with the fluctuation exchange(FLEX)
approximation. We find that a strong anisotropic band
renormalization is resulted from the Coulomb interaction with the
strongest effect occurring around the $\tilde{X}=(0,\pi)$ point,
which is defined in the unfolded Brillouin zone(BZ), and this
renormalization increases rapidly with the increase of the Hund's
coupling $J$ when $J>0.18U$($U$ is the intra-orbital Coulomb
interaction). Due to the band renormalization, the Fermi level for
the undoped case is slightly below the flat band centered around the
$\tilde{X}$ point which is the bottom of the band along the
$\tilde{X}$ to $\tilde{M}=(\pi,\pi)$ direction. As a result, for the
hole doped case, the Fermi level will situate below the flat band
along the $\tilde{X}$ to $\tilde{M}$ direction, though it still
crosses the renormalized band along the $\tilde{\Gamma}$ to
$\tilde{X}$ direction. In this case, the FS around $\tilde{X}$
consists of disconnecting patches. However, in the electron-doped
case, the Fermi level is lifted to be above the flat band and the
circular-like FS is formed. This result provides a possible
explanation for the different FS topology observed in the electron
doped and hole doped materials. We also carry out the same
calculation based on the two-orbital model~\cite{Raghu}, no similar
FS reconstruction has been found. This difference is ascribed to be
due to the orbital dependent renormalization. On the other hand, the
most favored pairing state mediated by the inter-orbital spin
fluctuations is found to be the extended $s$-wave with a sign change
between the electron and hole Fermi pockets, which is consistent
with the result obtained in the two-orbital
model~\cite{Yao,LeeDH,ChenWQ}. This indicates that the two models
share the similar physics as far as the pairing symmetry is
concerned, but exhibits difference in the band renormalization.

The paper is organized as following. In Sec.II, we present the
three-orbital model and introduce the FLEX method. In Sec.III, the
numerical results for the band renormalization is presented and
discussed. We also give a brief discussion on the pairing symmetry
in this section. In Sec.IV, we give a summary of the results.

\section{MODEL AND FLEX METHOD}

The model Hamiltonian consists of two parts,
\begin{eqnarray}
H=H_{0}+H_{int},
\end{eqnarray}
where the bare Hamiltonian $H_{0}$ is given by the three-orbital
model as introduced in Ref.~\cite{Lee}. In the unfolded(extended) BZ
for the reduced unit cell(only one Fe atom in the unit cell as in
Ref.~\cite{Mazin}), it can be written as
$H_{0}=\sum_{k}\Psi_{k}^{\dag}M_{k}\Psi_{k}$ with
\begin{eqnarray}
M_{k}=\left(\begin{array}{ccc}
        \varepsilon_{xz}(k+Q) & \varepsilon_{xz,yz}(k+Q) & \varepsilon_{xz,xy}(k) \\
        \varepsilon_{xz,yz}(k+Q) & \varepsilon_{yz}(k+Q) & \varepsilon_{yz,xy}(k) \\
        \varepsilon_{xz,xy}^\ast(k) & \varepsilon_{yz,xy}^\ast(k) &
        \varepsilon_{xy}(k)
      \end{array}\right),
\end{eqnarray}
and $\Psi_{k}=(c_{k+Q}^{xz},c_{k+Q}^{yz},c_{k}^{xy})^T$. Here the
diagonal elements of $M_{k}$ denote the dispersion of Fe-3d
orbitals $d_{xz}$, $d_{yz}$ and $d_{xy}$, while the others denote
the hybridization among them. Keeping up to the next nearest
neighbor hopping terms, we have
$\varepsilon_{xz}(k)=-\sum_{k}[2t_{1}\cos k_{x}+2t_{2}\cos
k_{y}+4t_{3}\cos k_{x}\cos k_{y}]$,
$\varepsilon_{yz}(k)=-\sum_{k}[2t_{1}\cos k_{y}+2t_{2}\cos
k_{x}+4t_{3}\cos k_{x}\cos k_{y}]$,
$\varepsilon_{xy}(k)=-\sum_{k}[2t_{4}(\cos k_{x}+\cos
k_{y})+4t_{5}\cos k_{x}\cos k_{y}]$,
$\varepsilon_{xz,yz}(k)=\varepsilon_{yz,xz}(k)=-\sum_{k}4t_{6}\sin
k_{x}\sin k_{y}$,
$\varepsilon_{xz,xy}(k)=\varepsilon_{xy,xz}^{\ast}(k)=-\sum_{k}2\mathrm{i}t_{7}\sin
k_{x}$~\cite{Lee},
$\varepsilon_{yz,xy}(k)=\varepsilon_{xy,yz}^{\ast}(k)=-\sum_{k}2\mathrm{i}t_{7}\sin
k_{y}$~\cite{Lee},
In order to reproduce the FS and band structure feature, we set the
parameters as $t_{1}=-1.0$($\approx0.4$eV), $t_{2}=0.7$,
$t_{3}=-0.8$, $t_{4}=-0.3$, $t_{5}=0.2$, $t_{6}=0.6$, $t_{7}=-0.35$.
As the three orbitals belong to the $t_{2g}$ manifold, we set the
same on-site energy to the three orbitals. In Fig.1, we show the
band structure and FS with $\mu=1.15$(electron density per site
$n=4.0$) corresponding to the parent compound. We find that this
three-orbital model can basically reproduce the main features of the
FS and band structure obtained in the LDA
calculation~\cite{Singh,Nekrasov}.
\begin{figure}
  \includegraphics[scale=0.7]{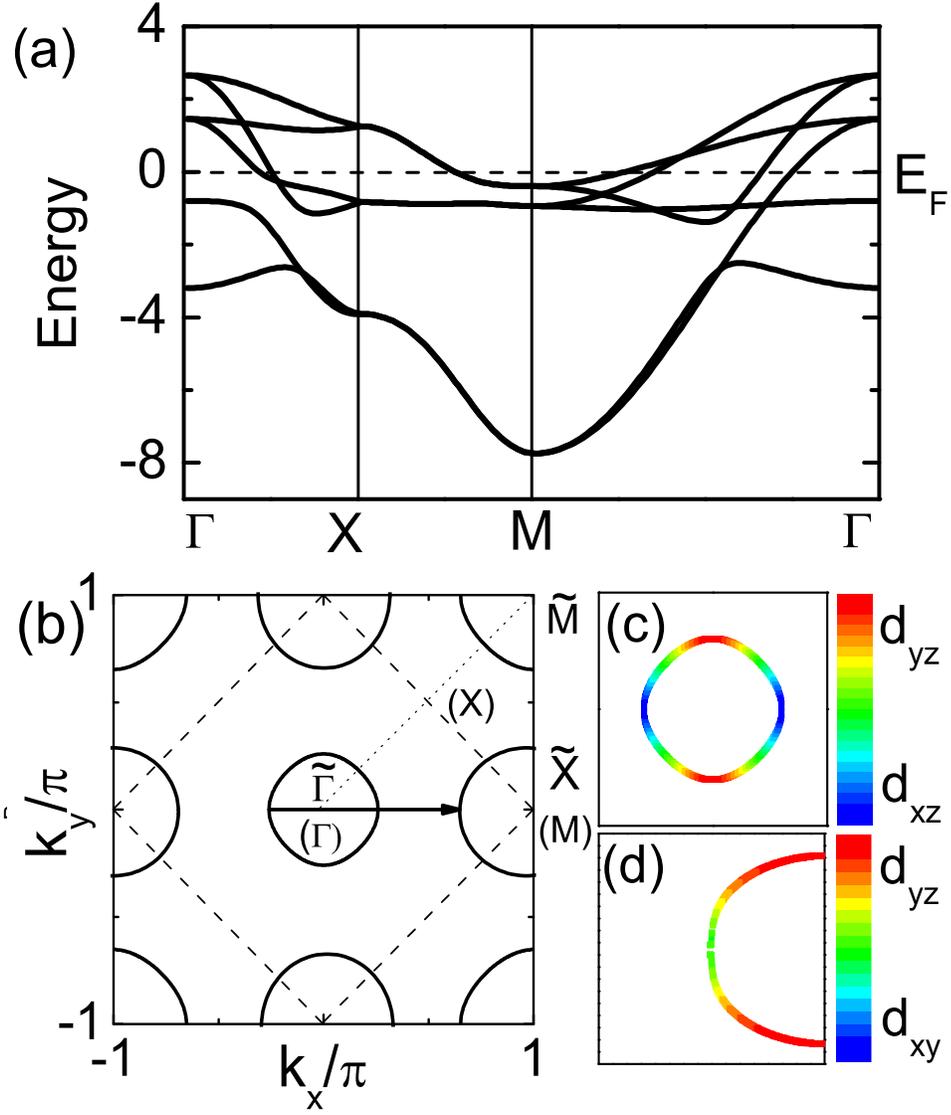}\\
  \caption{(color online) Band structure and FS of the three-orbital model.
  (a) The band structure in the folded BZ with $t_{1}=-1.0$,$t_{2}=0.7$,$t_{3}=-0.8$,$t_{4}=-0.3$,$t_{5}=0.2$,$t_{6}=0.6$,
  $t_{7}=0.35$ and $\mu = 1.15$. (b) The FS in the unfolded BZ.
  The dashed lines denote the boundary of the folded BZ.
  The line with an arrow denotes the nesting vector between the hole and electron Fermi pockets. Panels (c) and (d) replot the hole and
  the electron Fermi pockets shown in Fig.(b), respectively, with the different colors representing the weight of the different orbitals.}
\end{figure}
The interaction between electrons is included in $H_{int}$ as
following,
\begin{eqnarray}
H_{int}&=&\frac{1}{2}U\sum_{i,l,\sigma\neq\sigma'}c_{il\sigma}^{\dag}
c_{il\sigma'}^{\dag}c_{il\sigma'}c_{il\sigma} \nonumber \\
&+&\frac{1}{2}U'\sum_{i,l\neq
l^{'},\sigma,\sigma'}c_{il\sigma}^{\dag}
c_{il^{'}\sigma'}^{\dag}c_{il'\sigma'}c_{il\sigma}\nonumber \\
&+&\frac{1}{2}J\sum_{i,l\neq
l',\sigma,\sigma'}c_{il\sigma}^{\dag}c_{il'\sigma'}^{\dag}
c_{il\sigma'}c_{il'\sigma}\nonumber \\
&+&\frac{1}{2}J'\sum_{i,l\neq
l',\sigma\neq\sigma'}c_{il\sigma}^{\dag}c_{il\sigma'}^{\dag}
c_{il'\sigma'}c_{il'\sigma},
\end{eqnarray}
where $U$($U'$) is the intra-orbital(inter-orbital) Coulomb
interaction, $J$ the Hund's coupling and $J'$ the inter-orbital
pair hopping.

We carry out the investigation using the FLEX
approximation~\cite{flex}, in which the Green's function and
spin/charge fluctuations are determined self-consistently. For the
three-orbital model, the Green's function $\hat{G}$ and the
self-energy $\hat{\Sigma}$ are expressed in a $3\times3$-matrix
form, while the susceptibility $\hat{\chi}^{0}$ and the effective
interaction $\hat{V}$ have a $9\times9$-matrix form. The Green's
function satisfies the Dyson equation
$\hat{G}(k)^{-1}=\hat{G}^{0}(k)^{-1}-\hat{\Sigma}(k)$, where the
self-energy is given by
$\Sigma_{mn}(k)=\frac{T}{N}\sum_{q}\sum_{\mu\nu}V_{n\mu,
m\nu}(q)G_{\mu\nu}(k-q)$ and the bare Green's function reads
$\hat{G}^{0}(k)=(\mathrm{i}\omega_{n}-\hat{M}_{k}+\mu)^{-1}$. The
fluctuation exchange interaction is given by:
\begin{eqnarray}
V_{\mu
m,n\nu}(q)&=&\frac{1}{2}[3\hat{U}^{s}\hat{\chi}^{s}(q)\hat{U}^{s}
+\hat{U}^{c}\hat{\chi}^{c}(q)\hat{U}^{c}\nonumber\\
&&-\frac{1}{2}(\hat{U}^{s}+
\hat{U}^{c})\hat{\chi}^{0}(q)(\hat{U}^{s}+\hat{U}^{c})\nonumber\\
&&+3\hat{U}^{s}-\hat{U}^{c}]_{\mu m,n\nu},
\end{eqnarray}
with spin susceptibility
$\hat{\chi}^{s}(q)=[\hat{I}-\hat{\chi}^{0}(q)\hat{U}^{s}]^{-1}
\hat{\chi}^{0}(q)$ and charge susceptibility
$\hat{\chi}^{c}(q)=[\hat{I}+\hat{\chi}^{0}(q)\hat{U}^{c}]^{-1}
\hat{\chi}^{0}(q)$. The irreducible susceptibility is given by
$\chi^{0}_{\mu
m,n\nu}(q)=-\frac{T}{N}\sum_{k}G_{n\mu}(k+q)G_{m\nu}(k)$. In the
above, $T$ is temperature,
$k\equiv(\mathbf{k},\mathrm{i}\omega_{n})$ with $\omega_{n}=\pi
T(2n+1)$, and $\hat{I}$ the identity matrix. The interaction
matrix for the spin(charge) fluctuation
$\hat{U}^{s}$($\hat{U}^{c}$) is given by: For $i=j=k=l$,
$U^{s}_{ij,kl}=U$($U^{c}_{ij,kl}=U$); For $i=j\neq k=l$,
$U^{s}_{ij,kl}=J$($U^{c}_{ij,kl}=2U'-J$); For $i=k,j=l$ and $i\neq
j$, $U^{s}_{ij,kl}=U'$($U^{c}_{ij,kl}=-U'+2J$); For $i=l,j=k$ and
$i\neq j$, $U^{s}_{ij,kl}=J$($U^{c}_{ij,kl}=J$); For other cases,
$U^{s}_{ij,kl}=0$($U^{c}_{ij,kl}=0$).

After obtaining the renormalized Green's function $\hat{G}$, we can
solve the "Eliashberg" equation,
\begin{eqnarray}
\lambda\phi_{mn}(k)&=&-\frac{T}{N}\sum_{q}\sum_{\alpha\beta}\sum_{\mu\nu}
V^{s,t}_{\alpha m,n\beta}(q)G_{\alpha\mu}(k-q) \nonumber\\
& &\times G_{\beta\nu}(q-k)\phi_{\mu\nu}(k-q),
\end{eqnarray}
where the spin-singlet and spin-triplet pairing interactions
$\hat{V}^{s}$ and $\hat{V}^{t}$ are given by,
\begin{eqnarray}
\hat{V}^{s}(q)=\frac{3}{2}\hat{U}^{s}\hat{\chi}^{s}(q)\hat{U}^{s}-
\frac{1}{2}\hat{U}^{c}\hat{\chi}^{c}(q)\hat{U}^{c}+\frac{1}{2}
(\hat{U}^{s}+\hat{U}^{c}), \\
\hat{V}^{t}(q)=-\frac{1}{2}\hat{U}^{s}\hat{\chi}^{s}(q)\hat{U}^{s}-
\frac{1}{2}\hat{U}^{c}\hat{\chi}^{c}(q)\hat{U}^{c}+\frac{1}{2}
(\hat{U}^{s}+\hat{U}^{c}).
\end{eqnarray}
The most favorable SC pairing symmetry corresponds to the
eigenvector $\phi_{mn}(k)$ with the largest eigenvalue $\lambda$.

The Dyson equation, the self-energy and the interaction matrix
Eq.(4) form a closed set of equations and will be solved
numerically on $64\times64$ $\mathbf{k}$ meshes with 1024
Matsubara frequencies. By symmetry, we set $J'=J$ and use the
relation $U=U'+2J$. In the following calculation, the
intra-orbital Coulomb interaction $U=3.0$(about 0.3 total
bandwidth) is chosen, thus all interaction parameters are given by
giving the Hund's coupling $J$.

\section{RESULT AND DISCUSSION}

\subsection{Renormalization of band and Fermi surface}
\begin{figure}
  \includegraphics[scale=0.7]{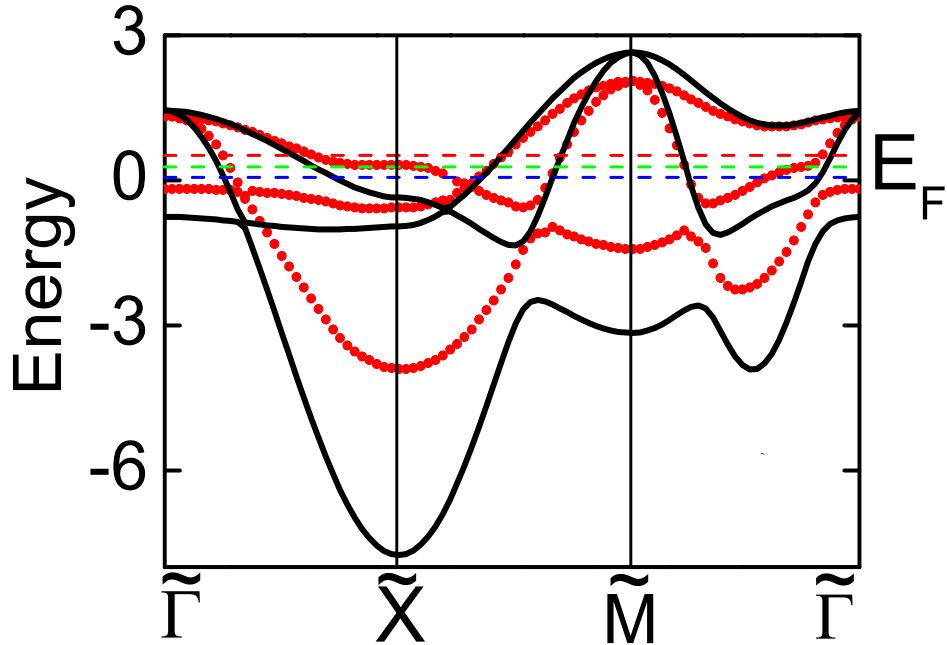}\\
  \caption{(color online) Renormalization of the energy band for $J=0.2U$, the red dotted(black solid)
  lines are the renormalized(bare) bands. The
  green, red and blue dashed lines indicate the Fermi levels for undoping, 10\% electron doping and
  20\% hole doping, respectively. To show the renormalization more clearly,
  the energy bands presented here are plotted in the unfolded BZ.}
\end{figure}
We show the renormalized bands for $J=0.2U$ (correspondingly
$U'=0.6U$ according to the relation $U=U'+2J$) together with the
bare bands in Fig.2, in which the renormalized bands are calculated
from the spectral function
$A(\mathbf{k},\omega)=-\frac{1}{\pi}\mathrm{Im}G(\mathbf{k},\omega)$
with $G(\mathbf{k},\omega)$ the analytic continuation of the
Matsubara Green's function $G(\mathbf{k},\mathrm{i}\omega_{n})$ by
the Pad$\acute{e}$ approximation. The Fermi levels depicted in Fig.2
are determined by calculating the number of electrons via the
renormalized Green's function, and the undoped case is determined
from the electron density per site $n=4.0$. It is clearly seen that
different bands exhibit different renormalization, and the strongest
renormalization occurs around the $\tilde{X}$ point. Compared to the
bare Fermi level, we find that the renormailzed Fermi level is
shifted up by 0.3($\approx0.12$eV) for the undoped case. These
features are consistent with the ARPES data~\cite{Lu,ding}. We
define a total bandwidth renormalization factor as $W_{B}/W_{R}$,
where $W_{B}$($W_{R}$) denotes the bandwidth of bare(renormalized)
band. The renormalization factors for different values of $J$ are
shown in Fig.3(a). It increases from 1.4 to 2.3 when $J$ is
increased to be around $0.5U$. In particular, a rapid rise is
observed when $J$ is larger than $0.18U$. This shows that the Hund's
coupling plays an important role to enhance the renormalization
effects.

\begin{figure}
  \includegraphics[scale=0.7]{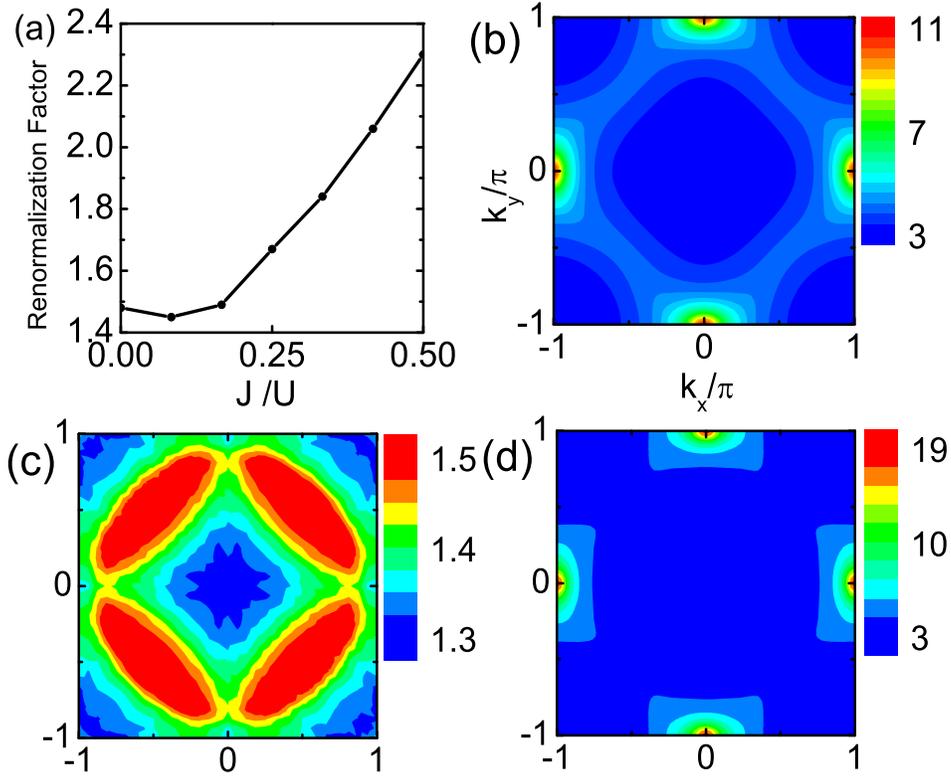}\\
  \caption{(color online) (a) Renormalization factor of total bandwidth for different values of $J$.
  (b),(c) and (d) are the spin susceptibilities for $J=0.2U$, $J=0.05U$ and $J=0.3U$ respectively.}
\end{figure}
For $U>U'$, the spin fluctuation will dominate over the charge
fluctuation. In Fig.3(b),we present the static spin susceptibility
$\chi^{s}({\bf q},\omega=0)=
\sum_{\mu\nu}\chi^{s}_{\mu\nu,\mu\nu}({\bf q},\omega=0)$ for
$J=0.2U$. It shows four peaks around $(0,\pm\pi)$ and $(\pm\pi,0)$
points, which is in agreement with the neutron scattering
experiments~\cite{Clarina}. This arises from the nesting between the
hole and the electron pockets connected with the vectors
$(0,\pm\pi)$ and $(\pm\pi,0)$, as shown in Fig.1(b). To show the
orbital contribution to the FS, we replot the hole and the electron
FS with different colors representing different weights of the three
orbitals in Fig.1(c) and (d), respectively. For the nesting part of
the FS, it is found that the $d_{yz}+d_{xy}$ orbitals contribute the
main weight to the electron pockets and the $d_{xz}$ orbital mainly
contributes to the hole pockets. This shows that the spin
fluctuation is mainly due to the inter-orbital particle-hole
excitations. As shown in Eq.(3), the Hund's coupling favors the
inter-orbital excitations, so the spin fluctuation around
$(0,\pm\pi)$ and $(\pm\pi,0)$ will be enhanced by the increase of
$J$. This is evidenced from the results shown in Fig.2(c) and 2(d),
where the spin susceptibilities for $J=0.05U$ and $J=0.3U$ are
presented, respectively. For a small $J$ $(0.05U)$, the $(0,\pm\pi)$
spin fluctuation disappears, while for a large $J$ $(J=0.3U)$ it is
enhanced. Because the band renormalization is increased with the
rise of $J$ as discussed above, it suggests that the strong
renormalization around the $\widetilde{X}$ point mainly comes from
the coupling to the spin fluctuation around $(0,\pm\pi)$.

\begin{figure}
  \includegraphics[scale=0.7]{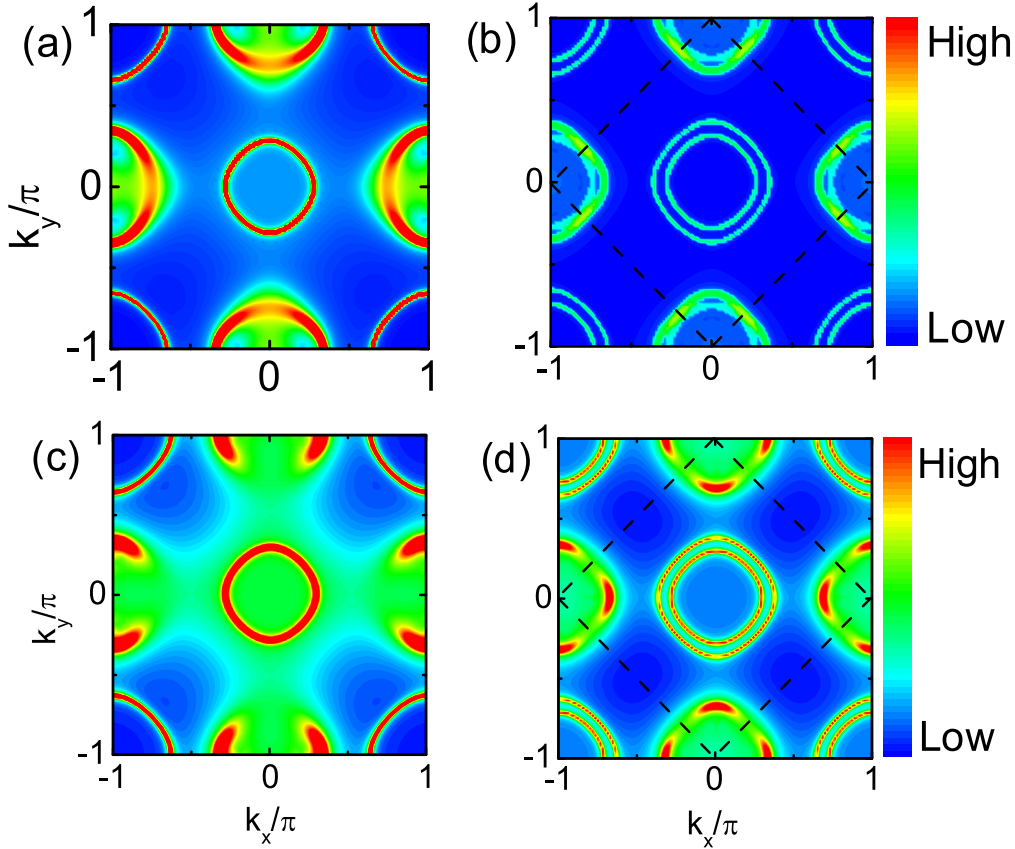}\\
  \caption{(color online) Renormalized FS for 10\% electron doping ((a) and (b)) and
  for 20\% hole doping ((c) and (d)). (a) and (c) are the results shown in the unfolded BZ, while (b) and (d) in the folded BZ.}
\end{figure}
Inspecting the details of the renormalized bands around the
$\widetilde{X}$ point in Fig.2, one will find that the energy band
near the Fermi level is renormalized strongly along the
$\widetilde{\Gamma}-\widetilde{X}$ direction, but less along the
$\widetilde{X}-\widetilde{M}$ direction. Another feature is that the
band crossing the Fermi level becomes flat around the
$\widetilde{X}$ point, and the flat portion situates at the bottom
of the band along the $\widetilde{\Gamma}-\widetilde{X}$ direction.
To identify the FS after renormalization, we calculate the
distribution of the spectral weight $A(\bf{k},\omega)$ integrated
over a narrow energy window $0.02(\approx8\mathrm{meV})$ around the
Fermi level, a method used usually in the ARPES experiments. The
results in the unfolded BZ are shown in Fig.4(a) for 10\% electron
doping and Fig.4(c) for 20\% hole doping. To compare with
experiments, their folded counterparts are shown in Fig.4(b) and
Fig.4(d), respectively. In the undoped case, the Fermi level(the
green dashed line) is just slightly below the flat band. Therefore,
for the electron doped case, in which the Fermi level is shifted
upwards as indicated by the red dashed line for the 10\% electron
doping, the Fermi level crosses the renormalized bands along both
the $\widetilde{\Gamma}-\widetilde{X}$ and the
$\widetilde{X}-\widetilde{M}$ directions. This gives rise to a large
complete electron Fermi pocket around the $\widetilde{X}$ point as
shown in Fig.2(a) in the unfolded BZ and in Fig.2(b) in the folded
one, which is similar to that predicted by the LDA
calculation~\cite{Singh,Nekrasov,Boeri} and observed in the ARPES
experiments~\cite{Lu,LiuC}. However, for the hole doped case, the
Fermi level(the blue dashed line for 20\% hole doping) is shifted
below the flat band, so that it does not cross the energy band along
the $\widetilde{\Gamma}-\widetilde{X}$ direction anymore.
Consequently, the complete electron Fermi pocket in the bare case is
broken and the spot-like portions are formed around the
$\widetilde{X}$ point, as clearly shown in Fig.4(c). After being
folded, the electron FS shown in Fig.4(d) reproduces the
experimental results in ARPES measurements~\cite{Liu,zhao}. We note
that due to the band renormalization the crossing between the two
energy band along the $\widetilde{X}-\widetilde{M}$ directions is
shifted up and approaches the Fermi level for 20\% hole doping. This
crossing gives rise to a large density of states. Therefore it adds
an enhancement in the spectral weight around the spot-like portions
and makes it be easily observed in experiments.

The anisotropic band renormalization along the
$\widetilde{\Gamma}-\widetilde{X}$ and the
$\widetilde{X}-\widetilde{M}$ direction can be traced to the
different orbital weights composing the energy bands along these two
directions. Specifically, the energy band near the $\widetilde{X}$
point along the $\widetilde{X}-\widetilde{M}$ direction is composed
of only the $d_{yz}$ orbital, while that along the
$\widetilde{\Gamma}-\widetilde{X}$ direction is a mix of the
$d_{xy}$ and $d_{yz}$ orbitals, as shown in Fig.1(c)and Fig.1(d) in
which the different colors represent the weight of the respective
orbitals. As noted above, the band renormalization is mainly due to
the scattering off the inter-orbital spin fluctuations composed of
the components $\chi_{xz,yz}$,$\chi_{xz,xy}$ and $\chi_{yz,xy}$. We
find that the magnitude of the components $\chi_{xz,xy}$ and
$\chi_{yz,xy}$ is the same, while it is larger than that of the
$\chi_{xz,yz}$. Therefore, the $d_{xy}$ orbital is renormalized most
strongly by the spin fluctuation. This suggests that the anisotropic
renormalization of the energy band near the $\widetilde{X}$ point is
orbital dependent.

\begin{figure}
  \includegraphics[scale=0.7]{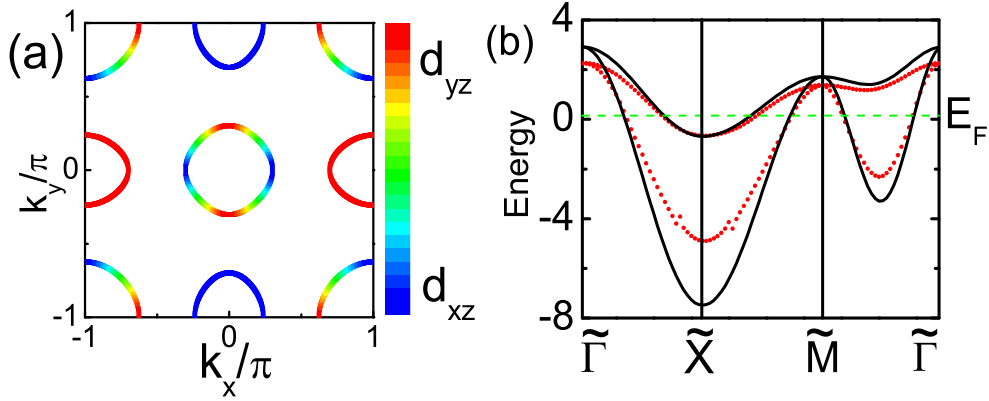}\\
  \caption{(color online) (a) Renormalization of the energy band for the two-orbital model,
  the red dotted(black solid) lines are the renormalized(bare) bands. The
  green line indicate the Fermi level for undoping. (b) The FS for the two-orbital model,
  the weight of $d_{xz}$ and $d_{yz}$ orbitals are depicted by different colors.}
\end{figure}

We note that in the two-orbital model~\cite{Raghu} the energy band
near the $\widetilde{X}$ point along both directions is composed
of one orbital $d_{yz}$ as shown in Fig.5(a), so the similar band
renormalization is expected. To show this, we have carried out the
same calculation for the two-orbital model and the result is
presented in Fig.5(b). Though the strong renormalization still
occurs in one of the bands around the $\widetilde{X}$ point, no
strong anisotropic renormalization between the
$\widetilde{\Gamma}-\widetilde{X}$ and the
$\widetilde{X}-\widetilde{M}$ direction can be seen. Thus, no
similar FS reconstruction as observed in the three-orbital model
and in experiments~\cite{Liu,zhao} can be obtained here.

\subsection{Symmetry of superconducting pairing}

\begin{figure}
  \includegraphics[scale=0.7]{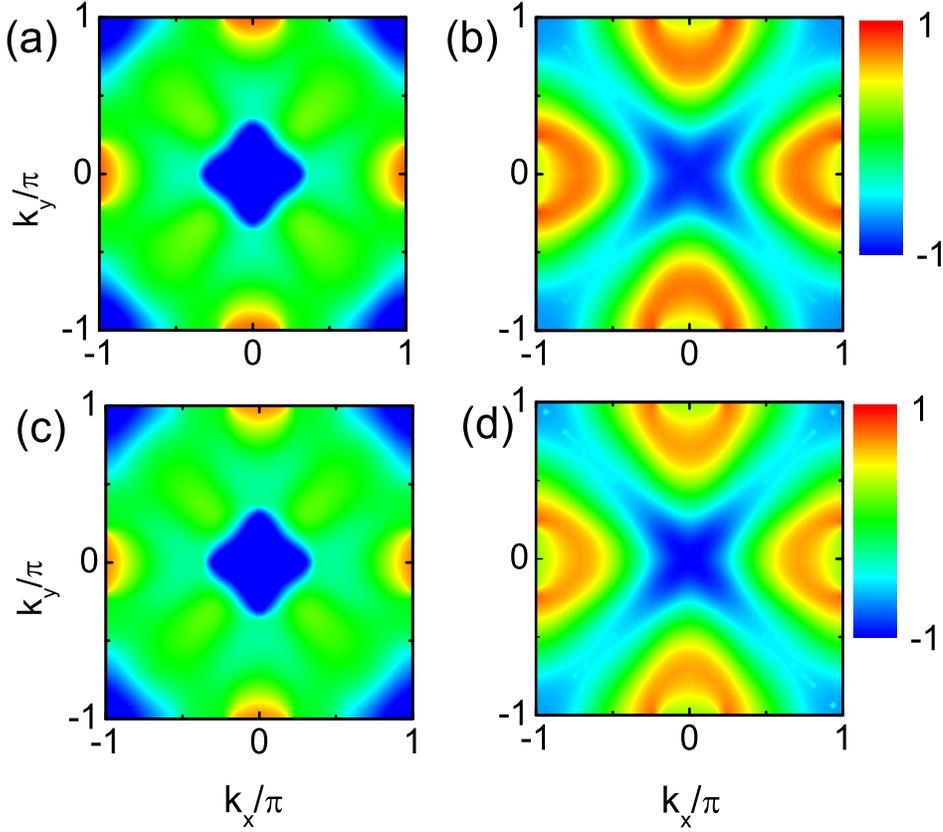}\\
  \caption{(color online) Pairing gap functions in the unfolded BZ for
  $U=3.0$ with $T=0.02$ for 10\% electron doping and 20\% hole doping.
  (a), (b), (c), (d) for $J=0.2U$ and (e), (f) for $J=0.05U$. (a) $\Delta_{hh}$ for electron doping.
  (b) $\Delta_{ee}$ for electron doping. (c) $\Delta_{hh}$ for hole doping.
  (d) $\Delta_{ee}$ for hole doping. (e) $\Delta_{hh}$ for electron
  doping. (f) $\Delta_{ee}$ for electron doping.}
\end{figure}

With the static spin susceptibility, we will further investigate the
pairing symmetry mediated by spin fluctuations. For $J=0.2U$, which
is the case with peaks around $(0,\pm\pi)$ and $(\pm\pi,0)$ points
in spin fluctuations, we find that the eigenvalue $\lambda$ has the
maximum value in the spin-singlet channel and it is nearly zero in
the spin-triplet channel at temperature $T=0.02$. We note that this
is not the case considered in Ref.~\cite{Lee}, where the
spin-triplet state is expected with a larger $J$ (such as $J>U/3$,
see also the analysis in Ref.~\cite{ChenWQ}). The obtained gap
functions of the hole band $\Delta_{hh}$ and the electron band
$\Delta_{ee}$ for 10\% electron doping and 20\% hole doping are
shown in Fig.3(a)-(d). It can be seen that the gap is basically an
extended $s$-wave, which is nodeless around all the Fermi pockets
and change sign between the hole pockets and the electron pockets.
This result is consistent with the experiment
measurements~\cite{zhao,ChenTY,Ding,Richard,Malone,MuG} and the
calculation based on the two-orbital model~\cite{Yao}. It is noted
that the pairing symmetry is similar for both the electron doped and
the hole doped systems, although their FS topology is different as
discussed above.

Because the spin fluctuation is dominant over the charge
fluctuation, the pairing interaction in the spin-singlet channel is
positive(repulsive) [See Eq.(6)] and strongest around the wave
vectors $\bf Q$ at which the spin fluctuation has a largest
intensity. For a repulsive pairing symmetry, the SC gap will satisfy
the condition $\Delta(\mathbf{k})\Delta(\mathbf{k}+\mathbf{Q})<0$ to
obtain the largest eigenvalue $\lambda$ of the "Eliashberg"
equation, as can be seen from Eq.(5). For $J>0.1U$, the inter-band
(inter-orbital) spin fluctuation is dominant and has peaks around
$(\pm\pi,0)$ and $(0,\pm\pi)$ which is the nesting wave vectors
connecting the hole and the electron Fermi pockets. Thus, the gap
will have an opposite sign between these two Fermi pockets. This
gives rise to the extended $s$-wave. Because the SC pairing is
mediated by the spin fluctuation around $(\pm\pi,0)$ and
$(0,\pm\pi)$, which is the same for the three-orbital and the
two-orbital models(see Ref.~\cite{Yao}), the same pairing symmetry
will be obtained based on these two models.

\section{CONCLUSION}

In a summary, we have investigated the band renormalization, FS
reconstruction and the symmetry of the supercondcuting gap in
iron-based superconductors using a three-orbital model. A strong
anisotropic band renormalization due to the scattering off the
inter-orbital spin fluctuations is found. The band renormalization
is shown to be orbital dependent. As a result, the electron Fermi
pocket exhibits different topology between the electron doped and
the hole doped cases, which provides a natural explanation for the
recent ARPES experiments. In addition, we have found that the most
favorable superconducting pairing symmetry mediated by the
inter-band (inter-orbital) spin fluctuations is the extended
$s$-wave.

\begin{acknowledgments}
We thank Z. Fang, X.J. Zhou, Q.H. Wang, Z.J. Yao and H.M. Jiang
for many helpful discussions. The work was supported by the NSFC
(10525415), the 973 project (2006CB601002,2006CB921800).
\end{acknowledgments}

\end{document}